\documentclass[12pt, draftclsnofoot, onecolumn]{IEEEtran}
\usepackage{graphicx}
\usepackage{amsthm}
\usepackage{epsfig}
\usepackage{latexsym}
\usepackage{amsfonts}
\usepackage{here}
\usepackage{rawfonts}
\usepackage[latin1]{inputenc}
\usepackage[T1]{fontenc}
\usepackage{calc}
\usepackage{capitalgreekitalic}
\usepackage{url}
\usepackage{enumerate}
\usepackage{color}
\usepackage[tbtags]{amsmath}
\usepackage{amssymb}
\usepackage{upref}
\usepackage{epic,eepic}
\usepackage{times}
\usepackage{dsfont}
\usepackage{comment}
\usepackage{cite}
\usepackage{bbm} 
\usepackage{multirow}

\usepackage{dsfont}

\newcounter{step}
\newlength{\totlinewidth}
  {\end{list}%
  \rule{\linewidth}{1pt}}
\newcounter{substep}

  {\end{list}}

\newlength{\aligntop}
\newlength{\alignbot}
 

\IEEEoverridecommandlockouts

\usepackage{algorithm}
\usepackage{algorithmic}

\makeatletter
\newcommand\semihuge{\@setfontsize\semihuge{19.3}{25}}
\makeatother

\makeatletter
\newcommand\semismall{\@setfontsize\semihuge{12.4}{15}}
\makeatother

\usepackage{subfigure}

\begin{document}

\title{\Huge Wireless Communications for Collaborative Federated Learning\vspace*{-0em}}

\author{{Mingzhe Chen}, \emph{Member, IEEE}, H. Vincent Poor, \emph{Fellow, IEEE}, Walid Saad, \emph{Fellow, IEEE}, and Shuguang Cui, \emph{Fellow, IEEE} \vspace*{-2em}\\ 

\thanks{M. Chen is with the Department of Electrical Engineering, Princeton University, Princeton, NJ, 08544, USA, and also with the Shenzhen Research Institute of Big Data, the Chinese University of Hong Kong, Shenzhen, 518172, China, Email: \protect{mingzhec@princeton.edu}.}
\thanks{H. V. Poor is with the Department of Electrical Engineering, Princeton University, Princeton, NJ, 08544, USA, Email: \protect\url{poor@princeton.edu}.}
\thanks{W. Saad is with the Wireless@VT, Bradley Department of Electrical and Computer Engineering, Virginia Tech, Blacksburg, VA, 24060, USA, Email: \protect{walids@vt.edu}.}
\thanks{S. Cui is with the Shenzhen Research Institute of Big Data and School of Science and Engineering, the Chinese University of Hong Kong, Shenzhen, 518172, China, Email: \protect{robert.cui@gmail.com} }
 }

\maketitle
%

\begin{abstract}
Internet of Things (IoT) services will use machine learning tools to efficiently analyze various types of data collected by IoT devices for inference, autonomy, and control purposes. However,  due to resource constraints and privacy challenges, edge IoT devices may not be able to transmit their collected data to a central controller for training machine learning models. To overcome this challenge, federated learning (FL) has been proposed as a means for enabling edge devices to train a shared machine learning model without data exchanges thus reducing communication overhead and preserving data privacy. However, Google's seminal FL algorithm requires all devices to be directly connected with a central controller, which significantly limits its application scenarios. In this context, this paper introduces a novel FL framework, called \emph{collaborative FL} (CFL), which enables edge devices to implement FL with less reliance on a central controller. The fundamentals of this framework are developed and then, a number of communication techniques are proposed so as to improve the performance of CFL. To this end, an overview of centralized learning, Google's seminal FL, and CFL is first presented. For each type of learning, the basic architecture as well as its advantages, drawbacks, and usage conditions are introduced. Then, three CFL performance metrics are presented and a suite of communication techniques ranging from network formation, device scheduling, mobility management, and coding is introduced to optimize the performance of CFL. For each technique, future research opportunities are also discussed. In a nutshell, this article will showcase how the proposed CFL framework can be effectively implemented at the edge of large-scale wireless systems such as the Internet of Things.

\end{abstract}

\section{Introduction}
Machine learning (ML) is witnessing an unprecedented interest from the wireless community \cite{8755300} driven by recent breakthroughs in deep learning, the rise of smart devices, and the wide availability of data. ML use
cases for wireless networks range from data analysis and prediction to wireless environmental monitoring as well as network control and optimization. However, centralized ML requires edge devices to transmit their collected data to a central controller for learning. In practical deployments of ML, such as in Internet of Things (IoT) systems, due to privacy issues and stringent resource (e.g., bandwidth and transmit power) constraints, edge IoT devices may not be able or willing to share their collected data with other devices or a central controller. For example, a wearable device can collect medical data from a given user. However, the user may not be willing to share such private data with other users. To enable edge IoT devices to train a shared ML model without data exchange, federated learning was proposed by Google in \cite{47976}. 

Federated learning (FL) is a distributed implementation of ML using which IoT devices can perform on-device ML model training while only exchanging ML model parameters with a central controller to collaboratively find a shared optimal ML model.
Keeping the data at IoT devices not only preserves privacy but may also reduce network traffic congestion. Due to the unique features of FL, a number of existing works, as summarized in \cite{niknam2019federated,8994206,8865093,khan2019federated}, studied the use of FL for the optimization of wireless network performance. 

In practice, to implement FL over IoT networks, edge devices must repeatedly transmit their trained ML models to a central controller via wireless links for ML model update. Due to limited wireless resources such as bandwidth, in a system such as the IoT, only a subset of devices can use FL. Meanwhile, ML models that are transmitted from IoT devices to a central controller (e.g., a base station) are subject to errors and delays caused by the wireless channel which affects the learning performance. Therefore, it is necessary to consider the optimization of wireless networks to improve FL performance, as pointed out in \cite{8952884,chen2019joint,9014530}. This emerging ``\emph{communications for FL}'' research area is the key focus of this work.

Recently, a number of surveys and tutorials related to FL over wireless networks appeared in \cite{niknam2019federated,8994206,8865093,khan2019federated} and \cite{8970161}. First, the works in \cite{niknam2019federated,8994206,8865093,khan2019federated} looked at the use of FL for communications, rather than the impact of wireless networking on FL. Moreover, all prior works in \cite{niknam2019federated,8994206,8865093,khan2019federated} and \cite{8970161} focused on the original FL developed by Google in \cite{47976} (called \emph{original FL} hereinafter), which requires all edge IoT devices to transmit their ML models to a central controller for ML model update. Hence, these existing surveys did not consider the implementation of FL with less or even no reliance on the central controller. Furthermore, they did not analyze how to use wireless communication techniques to optimize the FL performance. 


The main contribution of this article is to introduce a novel FL framework, dubbed \emph{collaborative FL}, that combines collaborative learning \cite{elgabli2019gadmm} with federated learning so as to enable edge devices to engage in FL without connecting to a central controller. To introduce this new framework, we first provide a detailed overview on centralized learning (CL), original FL (OFL), and collaborative FL (CFL), and summarize their advantages, drawbacks, and usage in Section \ref{se:system}. Then, in Section \ref{se:metrics}, we introduce three important performance metrics to quantify the CFL performance over IoT systems. Then, in Section \ref{se:communication}, we introduce several important communication techniques ranging from network formation, device scheduling, mobility management, and coding to optimize the CFL performance metrics. For each communication technique, we introduce the motivation for optimizing the CFL performance and then present an illustrative example and future research opportunities. Conclusions are drawn in Section \ref{se:conclusion}. 

\section{Preliminaries and Overview}\label{se:system}

 \begin{table}
{\scriptsize
\centering
  \newcommand{\tabincell}[2]{\begin{tabular}{@{}#1@{}}#2\end{tabular}}
\renewcommand\arraystretch{1}
 \caption{
    \vspace*{-0.05em}Summary of the Advantages, Drawbacks, and Usage Conditions of ML over Wireless Networks.}\label{ta:1}\vspace*{-0.5em}
\centering
\begin{tabular}{|c|l|l|l|}
\hline
 \multicolumn{1}{|c|}{\multirow{2}{*}{\textbf{}}}&\multicolumn{1}{|c|}{\multirow{2}{*}{\textbf{Advantages}}}   &  \multicolumn{1}{|c|}{\multirow{2}{*}{\textbf{ Drawbacks}}} &  \multicolumn{1}{|c|}{\multirow{2}{*}{\textbf{ Usage Conditions}}} \\ 
 &&&\\
\hline
\multirow{6}{*}{\textbf{CL}}& $\bullet$ \multirow{1}{5.3cm}{Ability to find a globally optimal ML model.}& $\bullet$ \multirow{1}{5.1cm}{Private data must be shared with a centralized controller such as a BS or cloud.} &$\bullet$ \multirow{1}{4.8 cm}{Each device must be willing to share its private data.} \\
&$\bullet$ \multirow{1}{5.3cm}{Ample computational resources and energy available for ML training.}&   &\\
&&$\bullet$ \multirow{1}{5.1cm}{Significant overhead for data collection.}&\multirow{1}{4.8cm}{$\bullet$ All devices can transmit data to the BS.}\\ 
&$\bullet$ \multirow{1}{5.3cm}{Imperfect wireless transmission has a minor impact on ML model training.}& $\bullet$ \multirow{1}{5.1cm}{Difficult to implement for resource and energy-limited edge devices such as IoT devices.}&\\
&&& \\
&$\bullet$ \multirow{1}{5.3cm}{Better performance for ML models with non-convex functions compared to FL.}&&\\
&&$\bullet$ \multirow{1}{5.1cm}{Large delays due to long-range transmission to a remote cloud or BS.}& \\
&&& \\
\hline

\multirow{6}{*}{\textbf{OFL}}& \multirow{1}{*}{{$\bullet$ Privacy-preserving framework.}} &$\bullet$ \multirow{1}{5.1cm}{Imperfect wireless transmission affects the ML model training process.} &$\bullet$ \multirow{1}{4.8 cm}{All devices must be able to transmit FL model parameters to a controller or aggregator (e.g., a BS).} \\
&$\bullet$ \multirow{1}{5.3cm}{Devices can learn a common ML task in a distributed manner.}&&\\
&&$\bullet$ \multirow{1}{5.1cm}{Number of users (and their data) that can perform FL is limited.}  &\\ 
&$\bullet$ \multirow{1}{5.3cm}{Ability to train ML models at device level.} &&$\bullet$ \multirow{1}{4.8 cm}{All devices must be able to receive the FL model parameters from the BS.} \\ 
&&$\bullet$ \multirow{1}{4.8 cm}{All devices must have a direct and reliable wireless connection to the BS.} &\\
&&&$\bullet$ \multirow{1}{4.8cm}{Devices can locally train ML models (at the edge).} \\ 
&&&\\
\hline
\multirow{5}{*}{\textbf{CFL}}&$\bullet$  \multirow{1}{5.3cm}{{Privacy-preserving framework.}} &$\bullet$ \multirow{1}{5.1cm}{Imperfect wireless transmission affects the ML model training process.} &$\bullet$ \multirow{1}{4.8cm}{A reliable communication link can be formed between any two devices that need to use CFL.} \\
&$\bullet$  \multirow{1}{5.3cm}{Ability to include more training data samples for training compared to OFL.} &&\\
&& $\bullet$  \multirow{1}{5.1cm}{Lower convergence speed compared to OFL.} &\\
&$\bullet$ \multirow{1}{5.3cm}{Amenability for implementation in large-scale systems (e.g., IoT) because CFL can accommodate more devices in the FL process compared  to OFL.}&&$\bullet$ \multirow{1}{4.8 cm}{Each device can locally train its ML model and aggregate the local FL models received from its associated devices.} \\ 
&& $\bullet$ \multirow{1}{5.1cm}{The ML model of each device at convergence may be different since each device connects to a subset of devices.}& \\
&&&\\
&&& \\
\hline
\end{tabular}
 \vspace{-0.2cm}
}
\end{table}

In this section, we introduce the basic architectures and differences between CL, OFL, and CFL. 

\subsubsection{Centralized Learning}

\begin{figure*}[htbp]
  \centering
  {\subfigure[Architecture of CL]{\includegraphics[width=3.9cm]{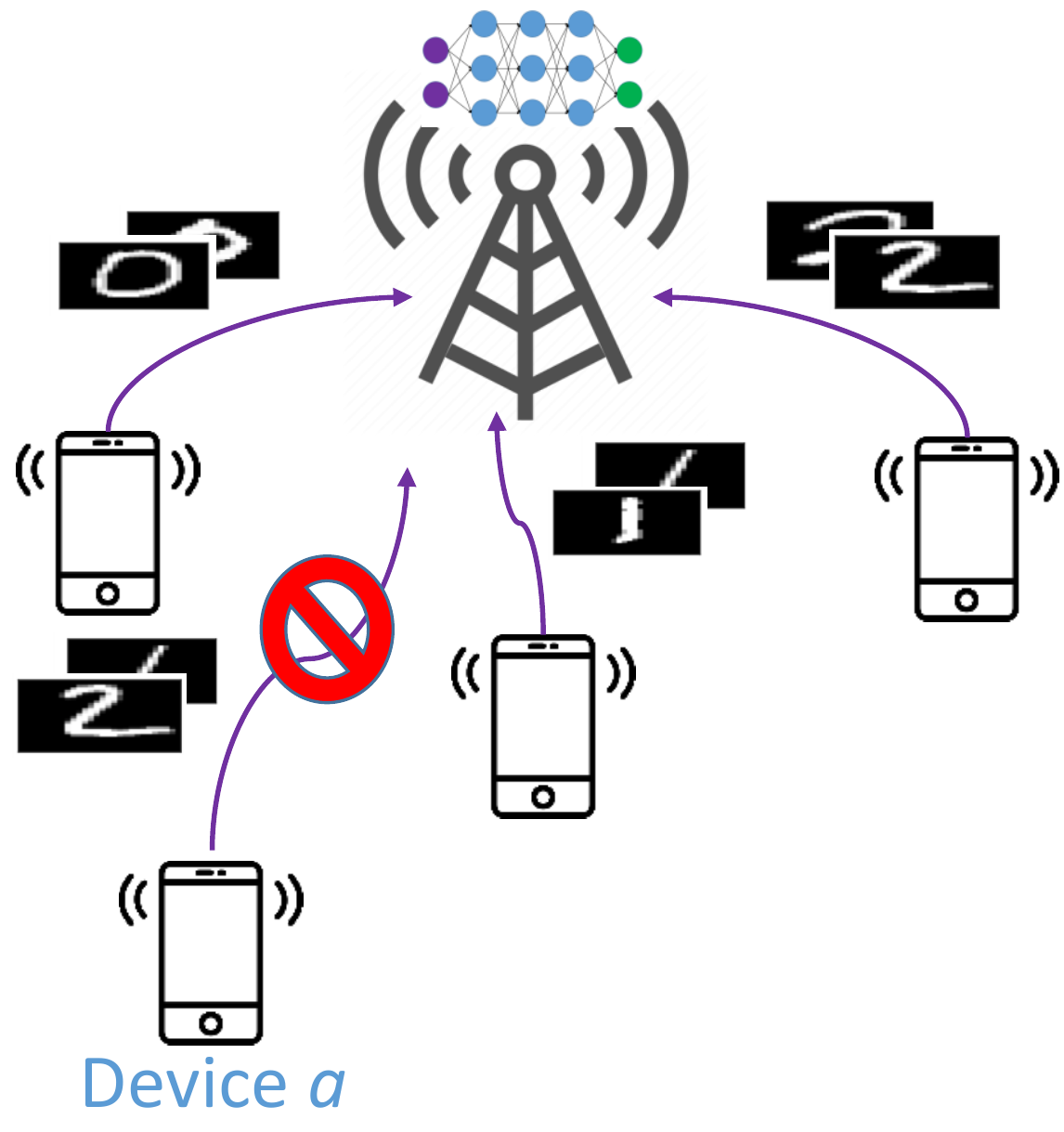}
\label{fig1a}}\hspace{0.65cm}
\subfigure[Architecture of OFL ]{\includegraphics[width=4.4cm]{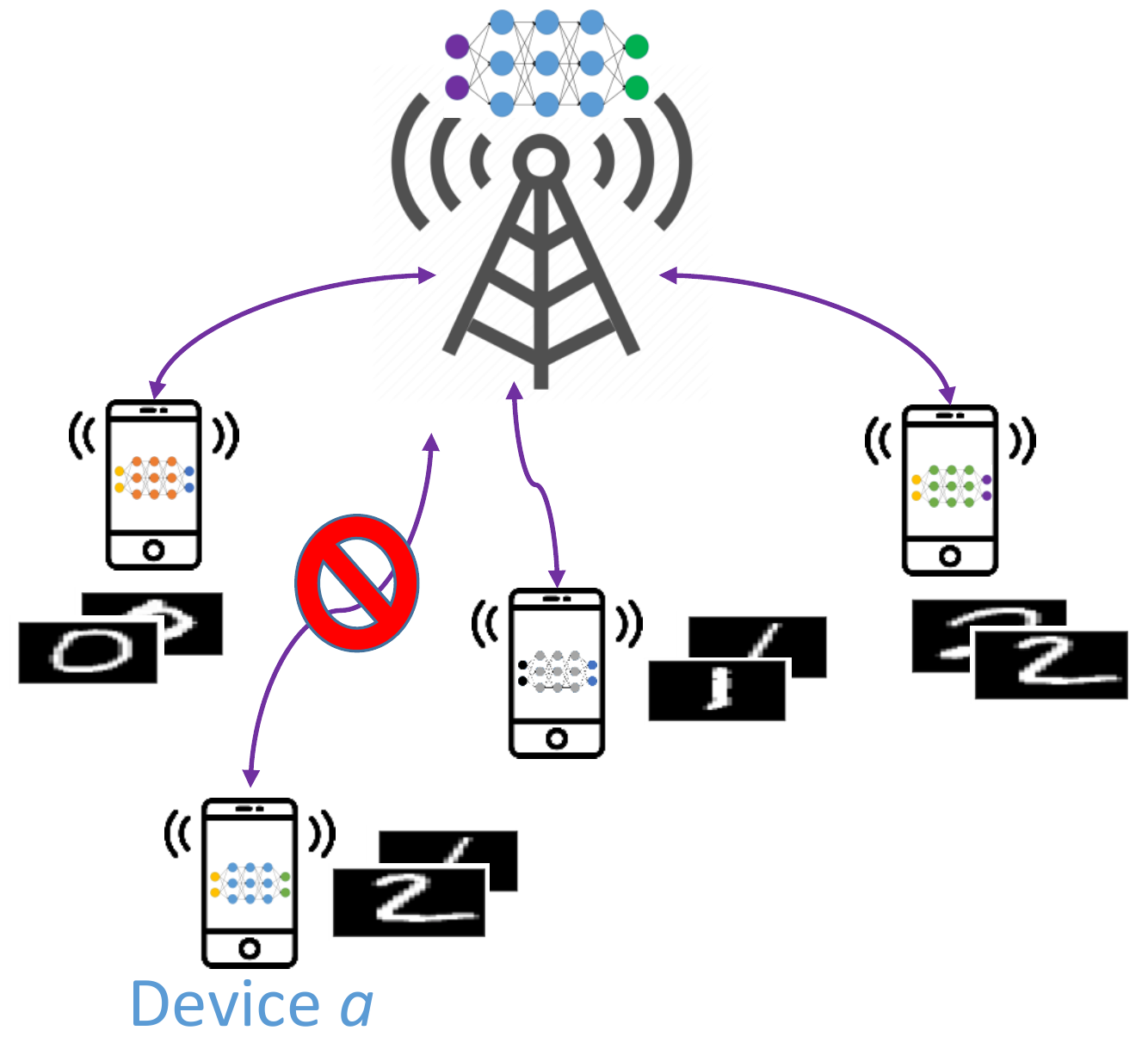}
\label{fig1b}}\hspace{0.65cm}
\subfigure[Architecture of CFL ]{\includegraphics[width=4.4cm]{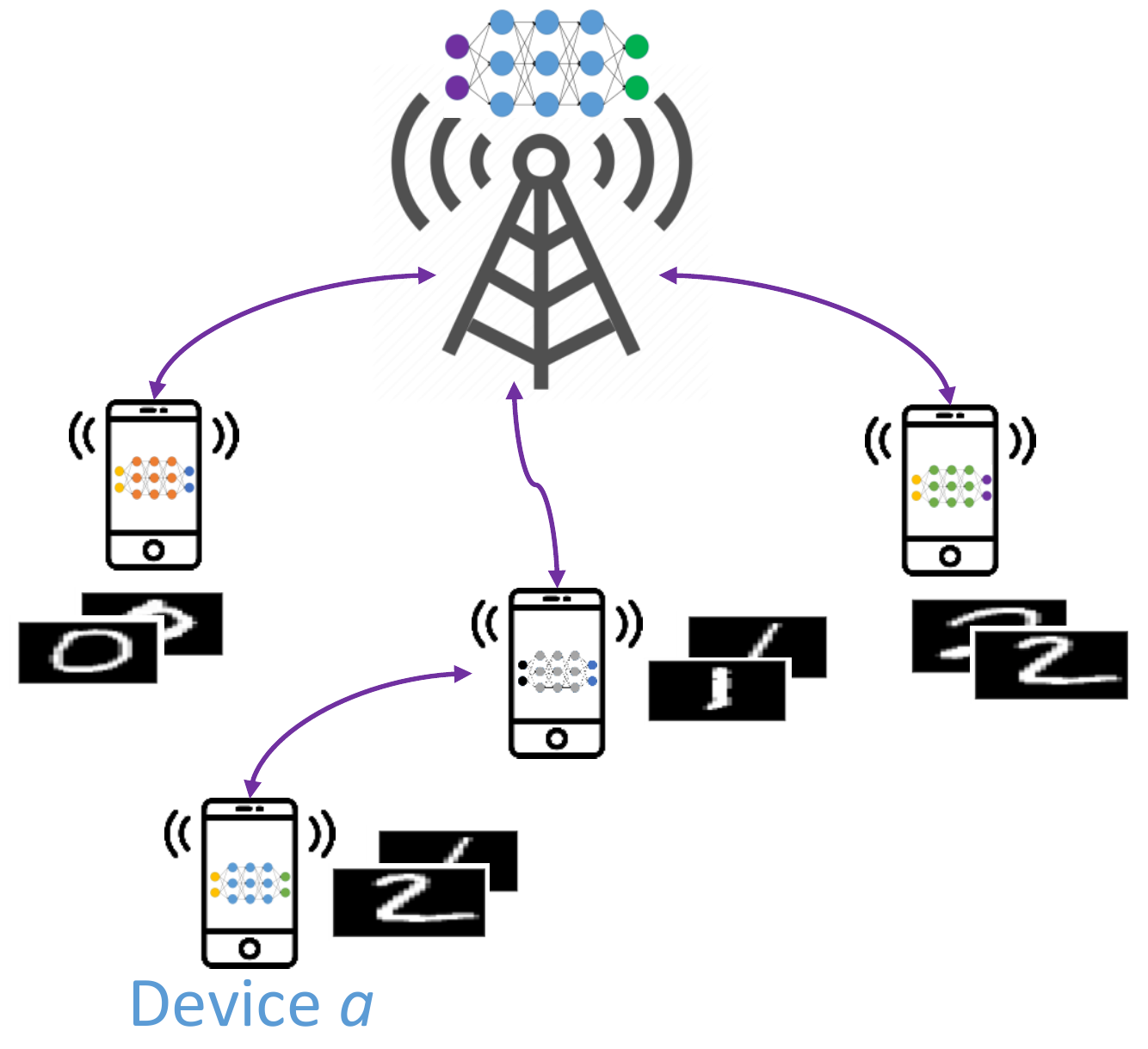}
\label{fig1c}} 
 }
  \caption{Architectures of centralized learning, original FL, and collaborative FL.}\label{fig1}
\vspace{-0.3cm}
\end{figure*}

As shown in Fig. \ref{fig1a}, CL needs only one ML model located at a base station (BS) or IoT cloud which works as a central controller. All devices must connect and send their data to the BS for training this ML model. Then, the BS will transmit the trained ML model to all devices. Hence, CL only requires the BS to communicate with all devices once so as to collect all devices' datasets.

Table \ref{ta:1} summarizes the advantages, disadvantages, and usage conditions of CL. The key advantage of CL is that it enables the BS or cloud to directly find a globally optimal ML model that minimizes the learning loss function value. Since the entire training process is completed by the BS, the ML training will not be affected by wireless network performance. However, imperfect wireless transmissions may introduce errors to the data used for training. Moreover, CL requires devices to transmit their collected data to the BS which leads to information leakage. In addition, significant overhead and resources are needed at the network and device levels to execute CL. 

\subsubsection{Original Federated Learning} To maintain privacy, Google's OFL framework allows each edge device to cooperatively train a shared ML model without data transmission. In OFL, both devices and the BS own an ML model with the same architecture, as shown in Fig. \ref{fig1b}. OFL is trained by an iterative learning process. First, all devices use their local data to train their local ML models and transmit their trained models to the BS. Then, the BS aggregates the received ML models, generated a new aggregate ML model, and transmits it back to all devices. Hereinafter, the ML model that is trained by an edge device is called \emph{local FL model} while the ML model generated by the BS is called \emph{global FL model}.
At convergence, the global FL model will be equal to all local FL models, which means that devices find a shared FL model and the lcoal FL model at convergence can be used to analyze all devices' datasets. 

The advantages, disadvantages, and conditions for use of OFL are summarized in Table \ref{ta:1}. The key advantage of OFL is that it preserves data privacy and can be implemented over devices with less overhead than centralized ML. However, OFL still requires all devices to transmit their local FL model parameters to a BS. Hence, imperfect and dynamic wireless transmission will significantly impact the convergence time and the performance of OFL.

\subsubsection{Collaborative Federated Learning} OFL requires all devices to send their local models to a BS, however, in practical IoT systems, devices may not be able to connect to the BS due to energy limitations or to a potentially high transmission delay. To overcome this challenge and facilitate the use of OFL in real-world IoT systems, we propose the concept of CFL using which devices can engage in FL without connecting to a BS or a cloud.


In CFL, devices that cannot connect to the BS directly can associate with neighboring users. 
For example, as shown in Fig. \ref{fig1b}, for OFL, device $a$ cannot connect to the BS and perform FL due to a potentially high transmission delay. However, in CFL,  as shown in Fig. \ref{fig1c}, device $a$ can connect to its closest device for performing FL. CFL is also trained iteratively.
First, each device transmits its trained local FL model to its connected devices or the BS. Then, the BS generates the global FL model and transmits it to the associated devices. Finally, each device updates its local FL model based on the local FL models received from other devices or the BS. In OFL, each device must train its local FL model using gradient descent (GD) methods while the BS aggregates the local FL models. However, in CFL, each device must both aggregate the local FL models received from other devices and train its local FL model. 

 \begin{figure*}[htbp]
  \centering
  {\subfigure[Simulation system]{\includegraphics[width=13cm]{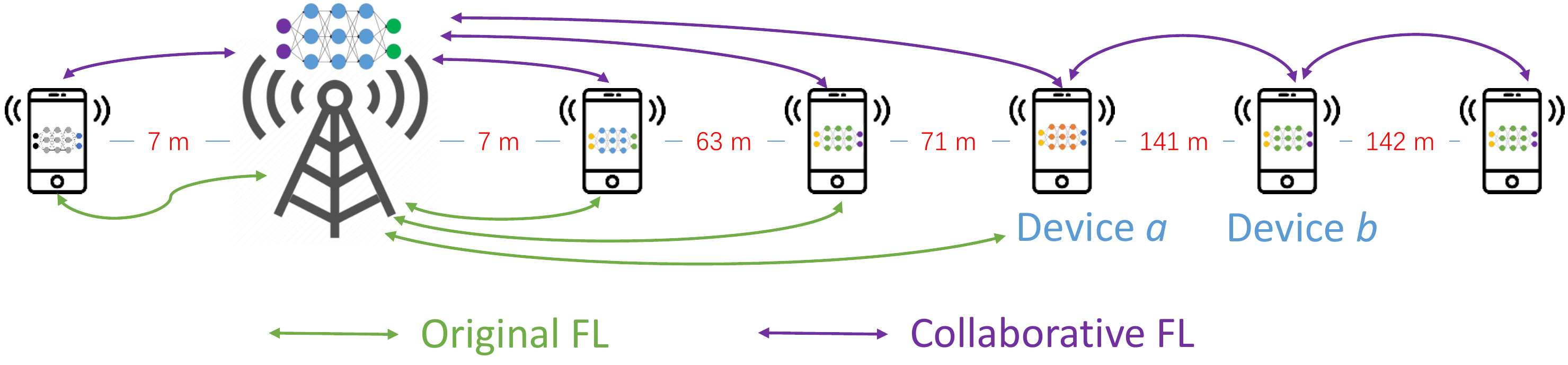}
\label{fig3a}}\hspace{0.1cm}
\subfigure[Simulation result]{\includegraphics[width=9cm]{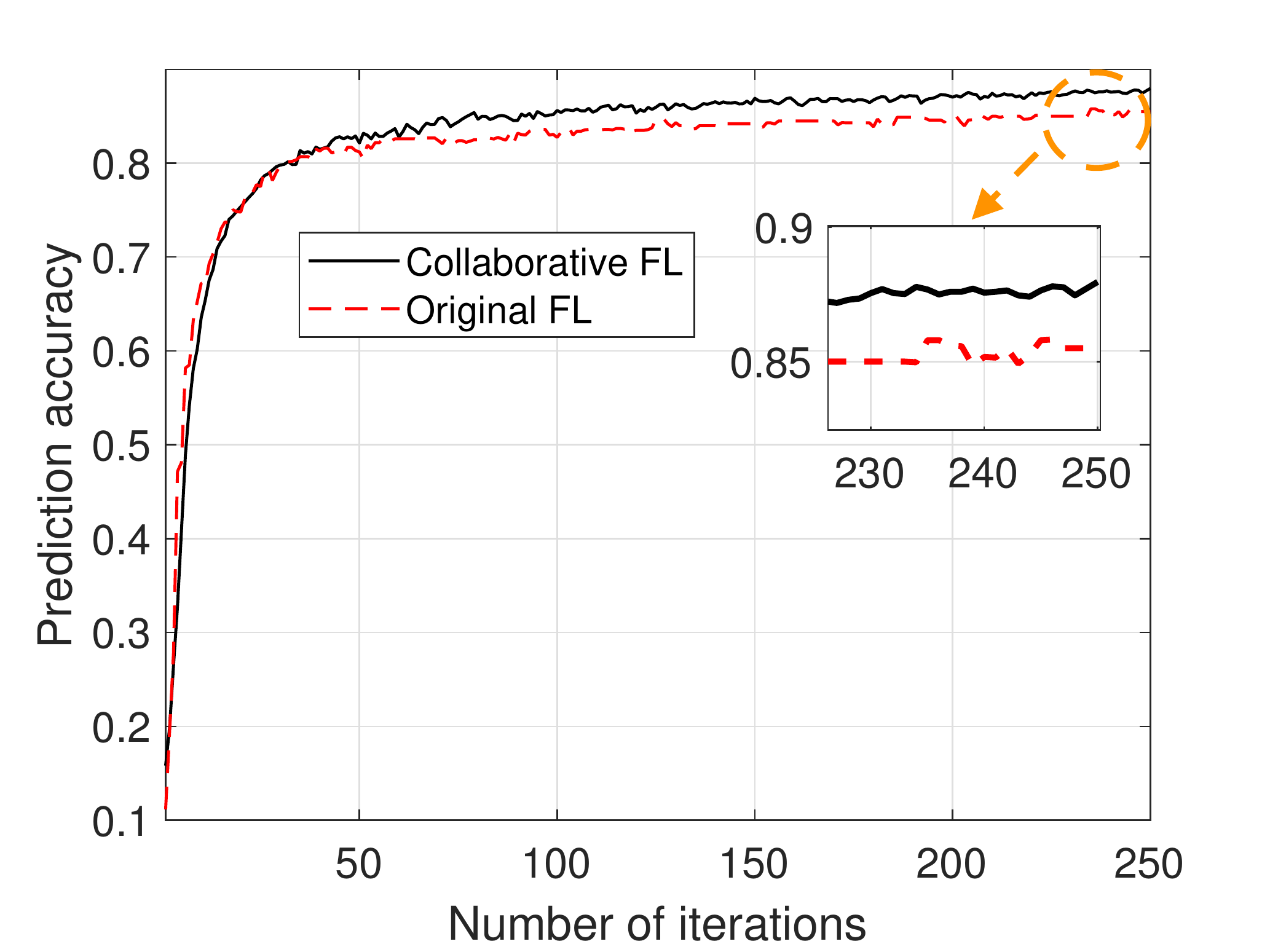}
\label{fig3b}} }
  \caption{Simulation system and result to show the performance of CFL and OFL. In this figure, a red digit is the distance between two adjacent devices.}\label{fig3}
\vspace{-0.3cm}
\end{figure*}

To show the difference between CFL and OFL, we implemented a preliminary simulation for a network having one BS and six devices, as shown in Fig. \ref{fig3a}. The local FL model of each device consists of a shallow feedforward neural network with 50 neurons. The MNIST  dataset \cite{MNIST} is used for training the local FL models at each device and each device has 500 data samples. OFL is used for comparison. The maximum time used for FL model parameter transmission is set to be 0.23 s. 


Fig. \ref{fig3b} shows how the identification accuracy changes over time. Fig. \ref{fig3b} demonstrates that CFL outperforms OFL. This is because, for OFL, only four devices can participate in FL and the other two devices have a delay larger than 0.23 s. Since CFL allows devices to connect to other devices and the transmission delay between any two neighboring devices is smaller than 0.23 s, six devices can participate.
In fact, CFL can also reduce the energy consumption for device $b$ since it only needs to transmit its ML model parameters to device $a$ instead of the BS.

Table \ref{ta:1} summarizes the advantages, disadvantages, and usage conditions of CFL. 
 The key advantage of CFL is that it 
enables the devices to perform the FL without transmitting local FL models to the BS, as shown in Fig. \ref{fig2}. 
Given the overview of CL, OFL, and CFL, we remark the following:

\begin{itemize}
\item Choosing between CL or FL depends on: a) willingness of data sharing, b) ML model data size, and c) size of the collected data of each device. For example, when devices agree to share the data and the size of the collected data is smaller than the ML model data size, CL is recommended.
 \item Choosing between OFL or CFL depends on: a) whether the BS performs FL and b) the connection and transmission delay between devices and the BS. For example, if all IoT devices need to implement FL without the BS, then CFL is more suitable.
\item OFL can be considered as a special case of CFL. In a network, if each device connects to all other devices, CFL is equivalent OFL. 
\end{itemize}


 \section{Performance of CFL over Wireless Networks}\label{se:metrics}
    \begin{table*}
 {\scriptsize
\centering
  \newcommand{\tabincell}[2]{\begin{tabular}{@{}#1@{}}#2\end{tabular}}
\renewcommand\arraystretch{1}
 \caption{
    \vspace*{-0.05em}Summary of the Wireless Factors that Affect the Performance Metrics and Suggested Solutions.}\label{ta:2}\vspace*{-0.5em}
\centering
\begin{tabular}{|c|l|l|l|}
\hline
 \multicolumn{1}{|c|}{\multirow{2}{*}{\textbf{}}}&\multicolumn{1}{|c|}{\multirow{2}{*}{\textbf{Wireless Factors}}}   &  \multicolumn{1}{|c|}{\multirow{2}{*}{\textbf{ Effects on FL}}} & \multicolumn{1}{|c|}{\multirow{2}{*}{\textbf{Suggested Solutions}}} \\ 
 &&&\\
\hline
\multirow{5}{*}{\textbf{Loss function value}}& \multirow{3}{*}{$\bullet$} \multirow{5}{3.8cm}{Limited wireless resources, e.g., bandwidth and computational resources.}&\multirow{3}{*}{$\bullet$} \multirow{5}{4cm}{Number of devices that can perform FL at each iteration is limited.}&$\bullet$ Probabilistic user scheduling. \\
& &&  $\bullet$ \multirow{1}{4.3cm}{Over the air techniques allowing devices to aggregate local FL models over wireless transmission.}  \\
& && \\
& && \\
& &&$\bullet$ Optimized network formation. \\
\cline{2-4}
& \multirow{2}{*}{$\bullet$ Limited transmit power.}& \multirow{2}{*}{$\bullet$ Errors in local FL models.}& $\bullet$ Channel coding and decoding. \\
&&& $\bullet$ Intelligent retransmission.\\
\hline

\multirow{7}{*}{\textbf{Convergence time}}& \multirow{2}{*}{$\bullet$} \multirow{4}{3.8cm}{Limited wireless resources, e.g., bandwidth, energy, and transmit power.}&$\bullet$ \multirow{1}{4cm}{Use of more time for local FL model parameter transmission.}&$\bullet$ Coding and decoding of FL model.  \\
&&&$\bullet$ FL model parameter prediction.\\
&&&$\bullet$ Over the air techniques. \\ 
& &&$\bullet$ Optimized network formation. \\
\cline{2-4}
&$\bullet$ \multirow{1}{*}{Limited computational resources.} &$\bullet$ \multirow{1}{4cm}{Number of local FL model updates at each CFL iteration is limited. }&$\bullet$ \multirow{1}{4cm}{Use of more global FL model updates.} \\ 
&&& \\ 
&&&$\bullet$ \multirow{1}{3.8cm}{Partial local FL model training. } \\ 
\hline
\multirow{4}{*}{\textbf{Energy consumption}}&  $\bullet$ \multirow{1}{3.8cm}{Limited wireless resources, e.g., bandwidth.} & $\bullet$ \multirow{1}{4cm}{Use of more energy for local FL model transmission.} &$\bullet$ Channel coding. \\
& &&$\bullet$ Optimized network formation.\\
&$\bullet$ \multirow{1}{*}{Wireless channel conditions.}&&$\bullet$  \multirow{1}{4cm}{Use of more local FL model updates.}\\
&&&\\
\hline
\multirow{5}{*}{\textbf{Reliability}}&  \multirow{5}{*}{{$\bullet$ Limited transmit power.}} & \multirow{5}{*}{$\bullet$ Errors in local FL models.} &$\bullet$ Channel coding. \\
& &&$\bullet$  \multirow{1}{4.2cm}{Improved device connection policy.}\\
&&&$\bullet$  \multirow{1}{4cm}{Use of more local FL model updates.}\\
&&&\\
& &&$\bullet$ Optimized network formation.\\
\hline
\end{tabular}
 \vspace{-0.2cm}
}
\end{table*}
 
We now introduce three key metrics for assessing the performance of CFL over wireless networks: a) value of the loss function, b) convergence time, and c) reliability.  
 \subsubsection{Loss Function Value}
An FL loss function is an objective function that devices try to minimize by adjusting their ML model parameters. For different learning tasks, the loss function will be different. 
 The loss function value is used to evaluate the performance of CFL. The CFL training purpose is to find an ML model that minimizes the loss function.
The FL loss function depends on the local FL models of all the participating devices. Hence, when those models are transmitted over wireless links, they experience transmission errors and delays which can negatively impact the loss function during training. Meanwhile, due to limited energy and computing resources, only a subset of devices can engage in CFL which decreases the total number of training data samples used for training the local FL models and increases the loss function value. Table \ref{se:metrics} summarizes the wireless factors that affect the FL loss function along with suggested solutions.
 

 \subsubsection{Convergence Time}\label{se:convergencetime}
 For CFL, the convergence time has three components: a) FL model parameter transmission delay, b) time needed by each device to train its local FL model, and c) number of iterations that FL needs to converge (i.e., the number of global FL model updates). The FL model parameter transmission delay depends on the data size of the FL model parameters and the data rate of the wireless link. The time used to train each device's local FL model depends on the FL model data size, the computational resources of each device, and the number of iterations (called number of local  FL model updates hereinafter) that each device uses to train its 
local FL model (using GD) at each FL iteration.
Note that as the number of local FL model updates increases, the number of global FL model updates decreases. The number of global FL model updates also depends on the limited spectrum resources that restrict the number of devices that engage in FL. Table \ref{ta:2} summarizes the wireless factors that affect the convergence time and the suggested solutions.
 
  \subsubsection{Energy Consumption} The energy consumption needed for training a CFL algorithm consists of four components: a) local FL model transmission, b) local FL model update, c) global FL model transmission, and d) global FL model aggregation. In particular, each device will spend energy for local FL model transmission and update while the BS needs to spend energy for global FL model transmission and aggregation. A tradeoff exists between the energy consumption of the local FL model update and the transmission energy. The energy consumption of CFL depends on the FL model data size, the distance between the BS and the devices, the convergence time requirement, and the target loss function value.  Table \ref{ta:2} summarizes the wireless factors that affect the energy consumption along with suggested solutions. 
  
  \subsubsection{Reliability} For CFL, we can define reliability as the probability that a CFL algorithm achieves a target FL loss function value.
At each CFL iteration, erroneous local FL models that are caused by imperfect wireless transmission must be abandoned by the devices. Hence, the number of local FL models used to generate the global FL model will decrease thus increasing the CFL convergence time and decreasing the loss function value.  Hence, a CFL algorithm may not be able to achieve a target FL loss function value due to imperfect wireless transmissions. Thus, the reliability of CFL depends on the wireless channel conditions. As the transmit power of each device increases, the number of erroneous local FL models decreases and thus increasing CFL reliability.
  Table \ref{ta:2} summarizes the wireless factors that affect the reliability and the suggested solutions. 
  
  \begin{figure*}[htbp]
  \centering
  {\subfigure[Grid topology]{\includegraphics[width=5cm]{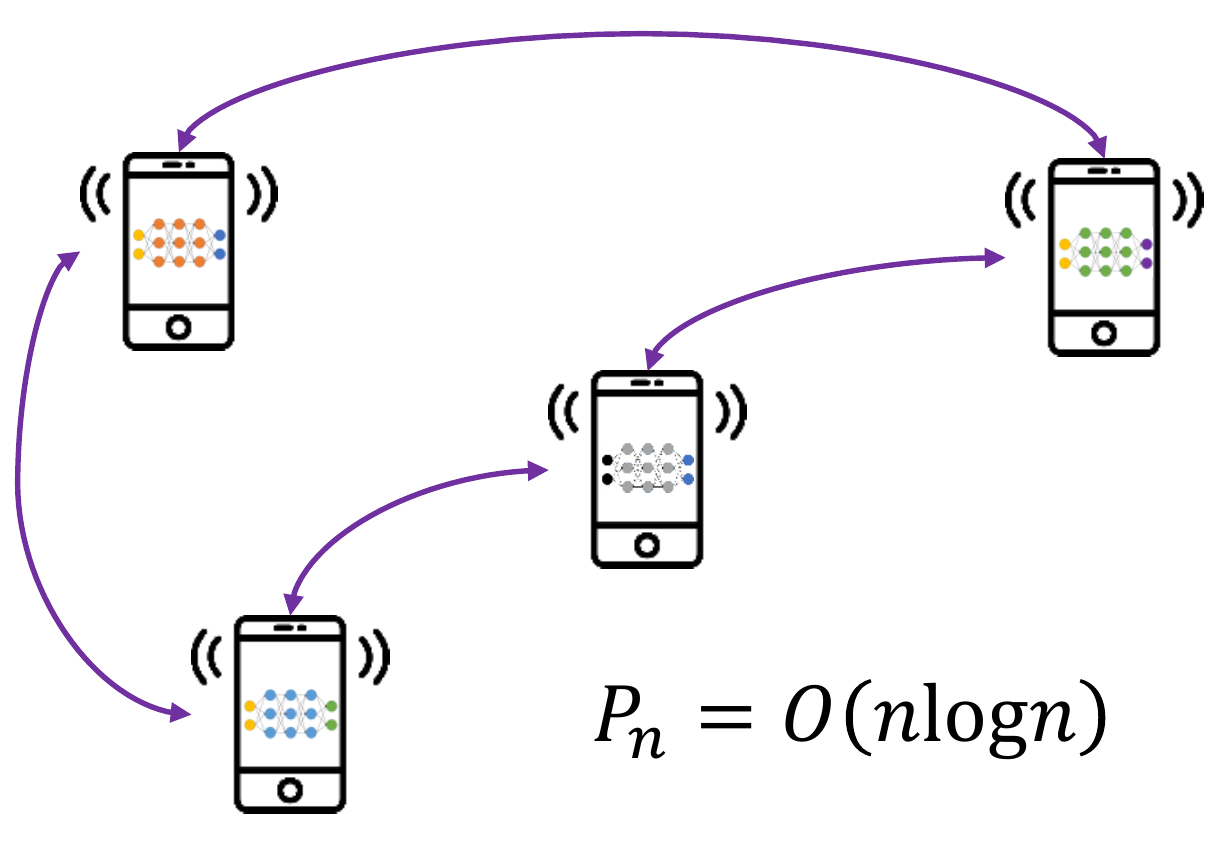}
\label{fig2a}}\hspace{0.12cm}
\subfigure[Path topology]{\includegraphics[width=5cm]{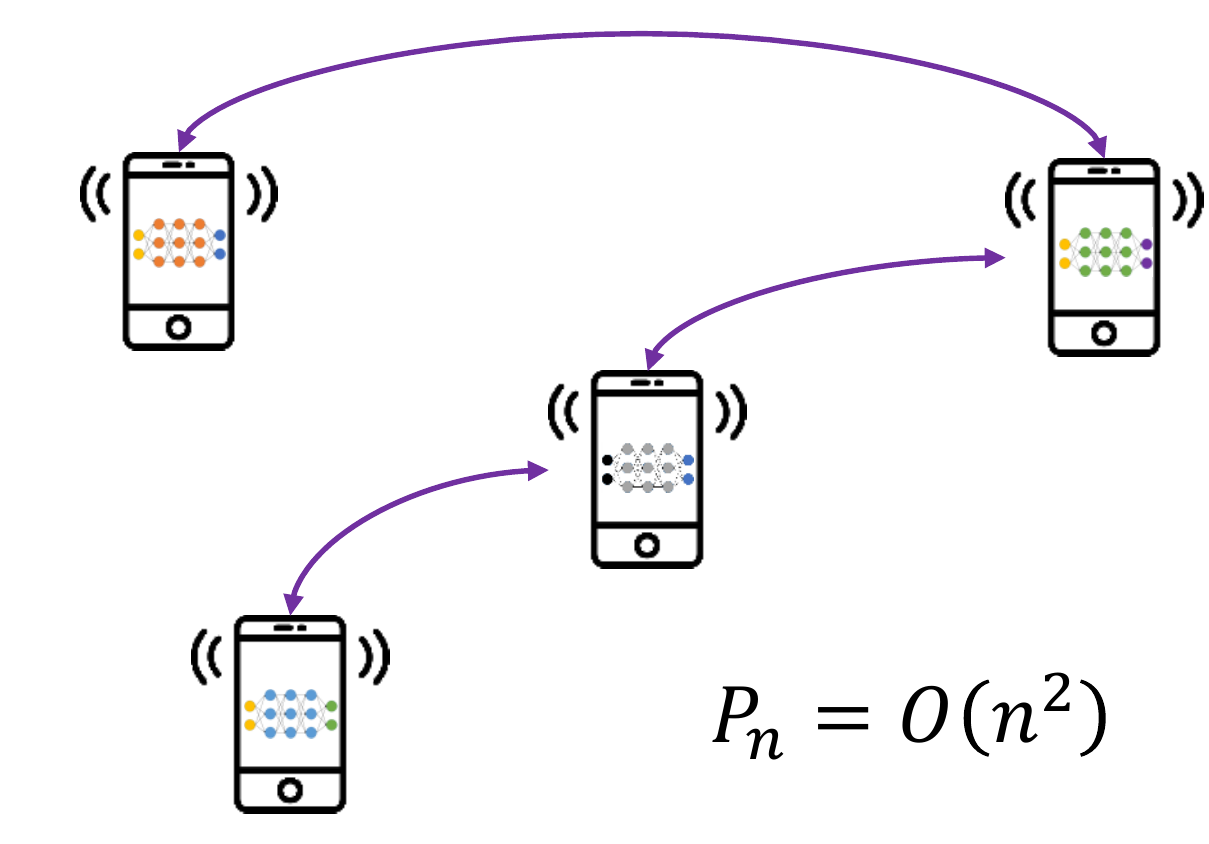}
\label{fig2b}}\hspace{0.12cm}
\subfigure[Complete topology]{\includegraphics[width=5cm]{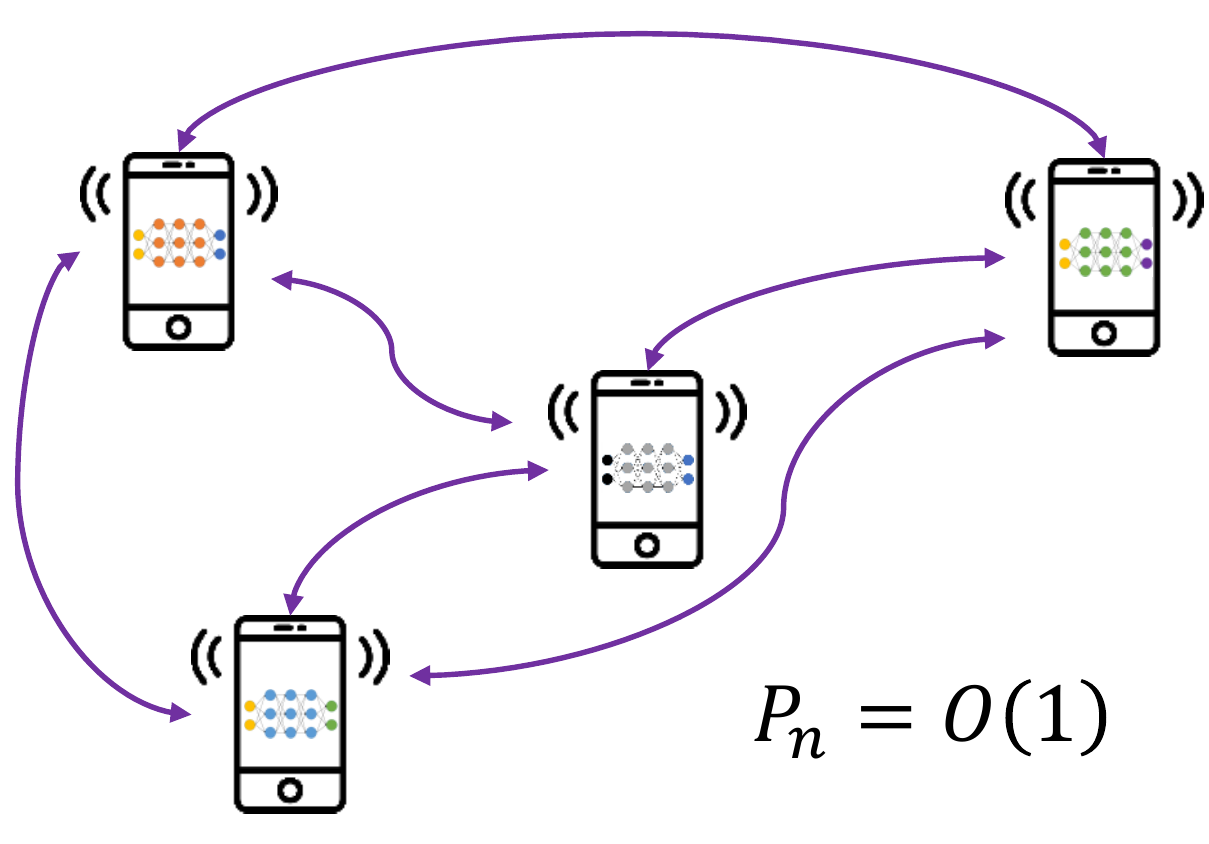}
\label{fig2c}} \hspace{0.12cm}
\subfigure[Star topology]{\includegraphics[width=5cm]{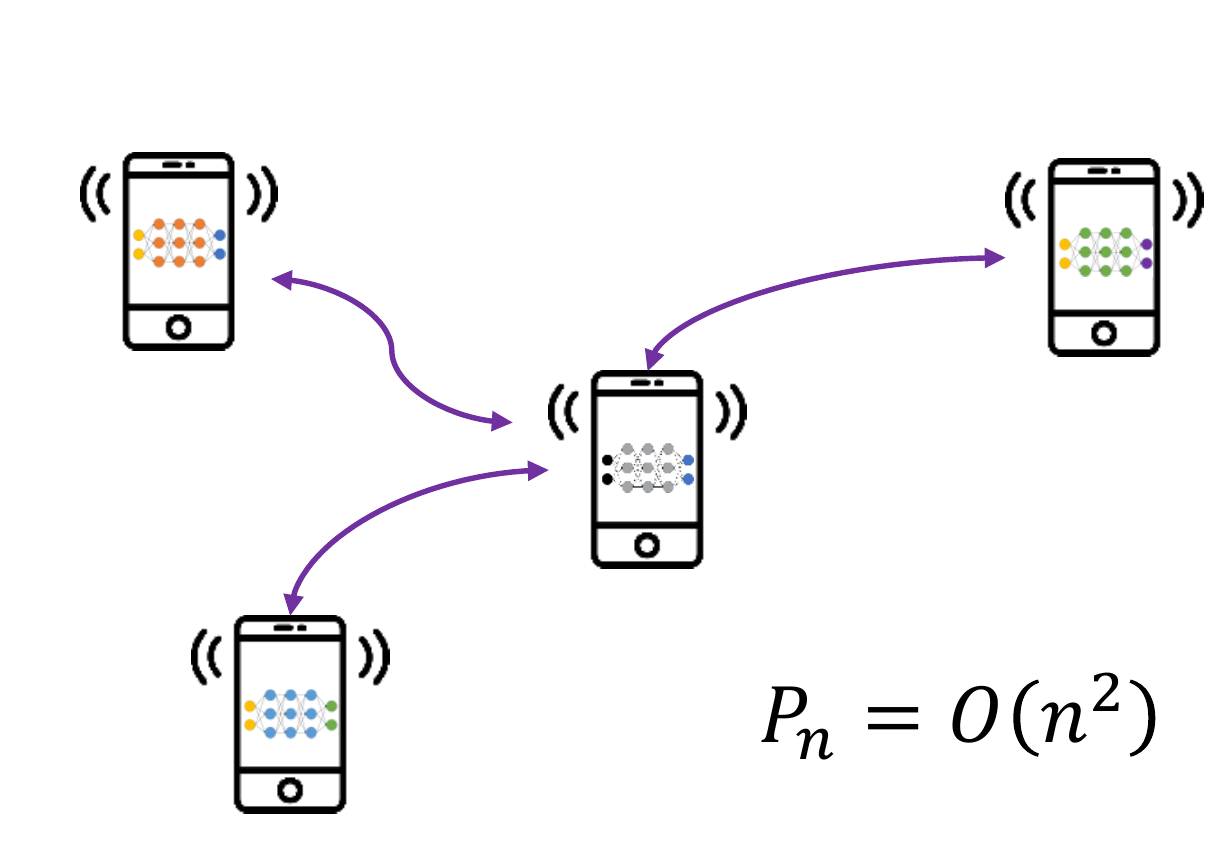}
\label{fig2d}}
 }
  \caption{Number of iterations needed to converge for different CFL algorithms with different topologies. In this figure, $O\left( {\frac{{\max \left( {{{\left( {{\boldsymbol{g}^0} - {\boldsymbol{w}^*}} \right)}^4},{L^4}P_n^2} \right)}}{{{\varepsilon ^2}}}} \right)$ is the upper bound of the number of iterations that a CFL algorithm needs to converge, where $n$ is the number of devices that perform the FL algorithm, $\varepsilon$ is the target accuracy which implies the difference between the optimal FL model and the FL model at convergence, $L$ is the upper bound of the gradient of the loss function, $\boldsymbol{g}^0=\frac{1}{n}\sum\nolimits_{i = 1}^n {\boldsymbol{w}_i^0}$ with $\boldsymbol{w}_i^0$ being the initial local FL model of device $i$, and $\boldsymbol{w}^*$ is the optimal local FL model at convergence.}\label{fig2}
\vspace{-0.3cm}
\end{figure*}

\section{Communication Techniques for Collaborative Federated Learning}\label{se:communication}
We now overview key techniques that can be used to improve the performance of CFL over wireless networks. 

\subsection{Network Formation}

The first fundamental step towards deploying CFL is to analyze the process of network formation using which devices can connect to one another to engage in a CFL task. In CFL, devices can form different network topologies. For example, IoT devices can form a grid topology for CFL, as shown in Fig. \ref{fig2a}. Naturally, the training complexity and the FL convergence time directly depend on the formed topology. Hence, for any given network scenario, it will be interesting to investigate the optimal CFL network topology using the metrics of Section \ref{se:metrics}.

Fig. \ref{fig2} shows the upper bound of the number of iterations for CFL convergence when assuming that the upper bound is derived based on the assumption that each device updates its local FL model using the Lazy Metropolis method and the GD method \cite{8340193}. 
Fig. \ref{fig2} shows that, when the number of links of each device increases, the number of iterations decreases because having more links increases the frequency of local FL model sharing.



Clearly, CFL yields interesting network formation research questions as follows:   
 \begin{itemize}
 \item \textbf{Optimal CFL network formation}: The optimal CFL network topology depends on the CFL performance metrics being optimized. Therefore, a fundamental CFL question is that of network formation: How can the devices interact to form an optimal network topology that maximizes the various CFL performance metrics and tradeoffs? To find the optimal CFL network topology, the first step is to define a proper utility function that jointly considers multiple dependent CFL performance metrics and network topology. Given the defined utility function, one must develop network formation algorithms to optimize the utility function. Both centralzied and distributed solutions can be developed. Centralized solutions such as searching based algorithms may be able to find the globally optimal network topology. However, the implementation of centralized solutions requires all devices' information such as locations or wireless channel conditions, which is impractical for a large-scale and dynamic IoT system. 
For distributed solutions, one can adopt a game-theoretic approach, particularly using network formation games \cite{han2012game}. In network formation games, each device is seen as an individual agent whose goal is to form a graph with neighboring devices so as to optimize the CFL performance metrics. The CFL performance (e.g., utility) depends on the entire graph and decision of all agents which makes the use of game theory suitable. One unique feature of the CFL network formation game is that it could be dynamic and requires far-sighted decision making. That is an angle that has only been studied in limited prior works as discussed in \cite{han2012game}.

\item \textbf{Network formation with asynchronous training}: Under asynchronous FL training, IoT devices will update and transmit their local FL models at different time slots. 
Due to limited computing and wireless resources, each device may not want to update its local FL model until it receives all local FL models of its associated devices. Using asynchronous training can increase local FL model update frequency and the data rate of each device which reduces the convergence time. In asynchronous training, the number of devices that need to transmit the local FL models is time-varying. Hence, the network topology must be adapted to the changes in the number of devices that must transmit local FL models.
Here, one must determine the frequency with which the network topology must be updated according to 
 the number of participating devices. Note that each network topology update will change the wireless resource allocation and device association schemes so as to improve CFL performance metrics such as convergence time. However, network topology updates will also introduce communication overhead such as network state information sharing.  



 \item \textbf{Network formation with partial network information}:
In actual IoT, each device may not completely
know the network architecture, device locations, and network composition.
 Due to this limited information, the number of devices that each device can connect to is limited and hence devices may not be able to form a network topology that satisfies the CFL usage conditions (see Table \ref{ta:1}). Therefore, there is a need to investigate a globally optimal network formation for IoT devices with partial information. Since most existing complexity results related to network formation (e.g., see  \cite{8340193}) assume that each device has complete information, they cannot be used for devices with partial network information. Meanwhile, due to partial network information, devices may form several unconnected small device groups. Hence, a multi-layer network formation must be designed. For example, in the first layer, devices will exchange their local FL model parameters in their own groups while the local FL model parameters are exchanged over multiple groups in the second layer. 
The designed scheme must balance the communication overheads and training complexity among multiple layers. 

\end{itemize}

\subsection{Device Scheduling} 
Due to energy constraints and wireless resource limitations, the number of devices that can engage in CFL is limited.
 Hence, an IoT device may update its local FL model using the local FL models of a subset of devices thus decreasing the CFL convergence time. Therefore, it is necessary to find an optimal device scheduling policy that can determine the frequency and which iterations that each device engages in CFL so as to optimize the CFL performance metrics. 


Device scheduling plays an important role in training CFL and it also faces several interesting research problems:
 \begin{itemize}
\item \textbf{Data importance-aware device scheduling}: In CFL, the contribution of each device's dataset on the update of a local FL model can be seen as the data importance of that device's dataset.
The data importance of each device depends on the number of training data samples and the data distribution. For instance, if a device has a large number of training data samples, its local FL model will be allocated a large weight within the local FL model update. Since only a subset of devices can perform FL at each iteration, it is necessary to design data importance-aware device scheduling policies for improving convergence speed. 
In particular, one must first build a data importance model that jointly considers the number of training data samples, data distribution, and data uniqueness.
Meanwhile, in CFL, devices cannot share data and, hence, each device may not be able to directly know the data importance of other devices. Therefore, there is a need to find a method to learn the data importance of other devices from their transmitted local FL model parameters. In addition, one must determine the frequency of local FL model update for devices with different data importance. Note that increasing the update frequency of the devices with high data importance can improve convergence speed but it also increases the loss function value.  




\item \textbf{Device scheduling for multiple FL tasks}: In a wireless network, a device may perform multiple FL algorithms simultaneously. Therefore, it will be interesting to design a device scheduling policy that enables devices to efficiently train multiple FL models and transmit the trained FL models to other devices simultaneously. Since each FL task has its specific convergence time requirement and target loss function value, the developed device scheduling policy must determine which FL model must be trained first and which FL model must be transmitted first so as to satisfy the requirements of each FL task. Moreover, since the convergence time of each FL task is different, the designed scheduling policy must be adapted to the changes in the number of incomplete FL tasks.

\item \textbf{Device scheduling and network formation for mobile devices}: In an IoT system, several devices, such as cars and drones, are mobile. The connections among different devices and the wireless network performance will change depending on the mobility of the devices thus affecting
the CFL performance. Meanwhile, device mobility will increase the frequency of devices changing their connections thus slowing down the CFL training process. Therefore, it is necessary to study device scheduling and network formation for mobile devices. 
In OFL, devices will transmit their local FL models to a static BS. However, in CFL, mobile devices must transmit their local FL models to other mobile devices. Hence, the devices' locations and connections are correlated in space (i.e., between two connected devices) and time (i.e., between time slots). For example, for two devices moving in parallel, although the location will be changing, the distance between the two devices remains constant. As a result, the change of their locations will not increase the local FL model transmission delay. Therefore, one must first build a model to capture the effect of spatio-temporal correlation of device locations and connections on the FL performance metrics. Then, it must investigate how to use spatio-temporal correlation to optimize device scheduling and network topology policies and the frequency of changing these policies.
 

\end{itemize}

 \begin{figure*}[htbp]
  \centering
  {\subfigure[CFL scenario used for our simulations.]{\includegraphics[width=12cm]{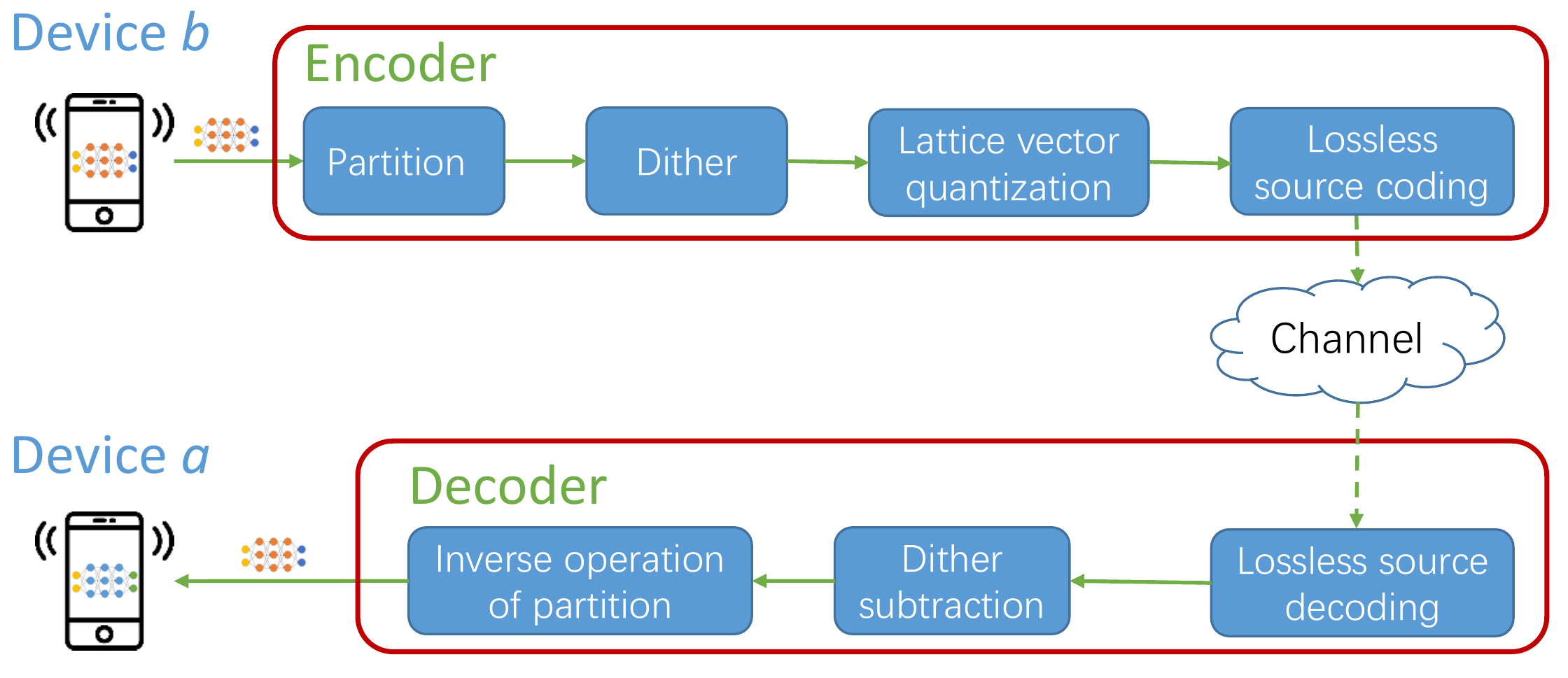}
\label{fig4a}}\hspace{0.1cm}
\subfigure[Simulation result]{\includegraphics[width=9cm]{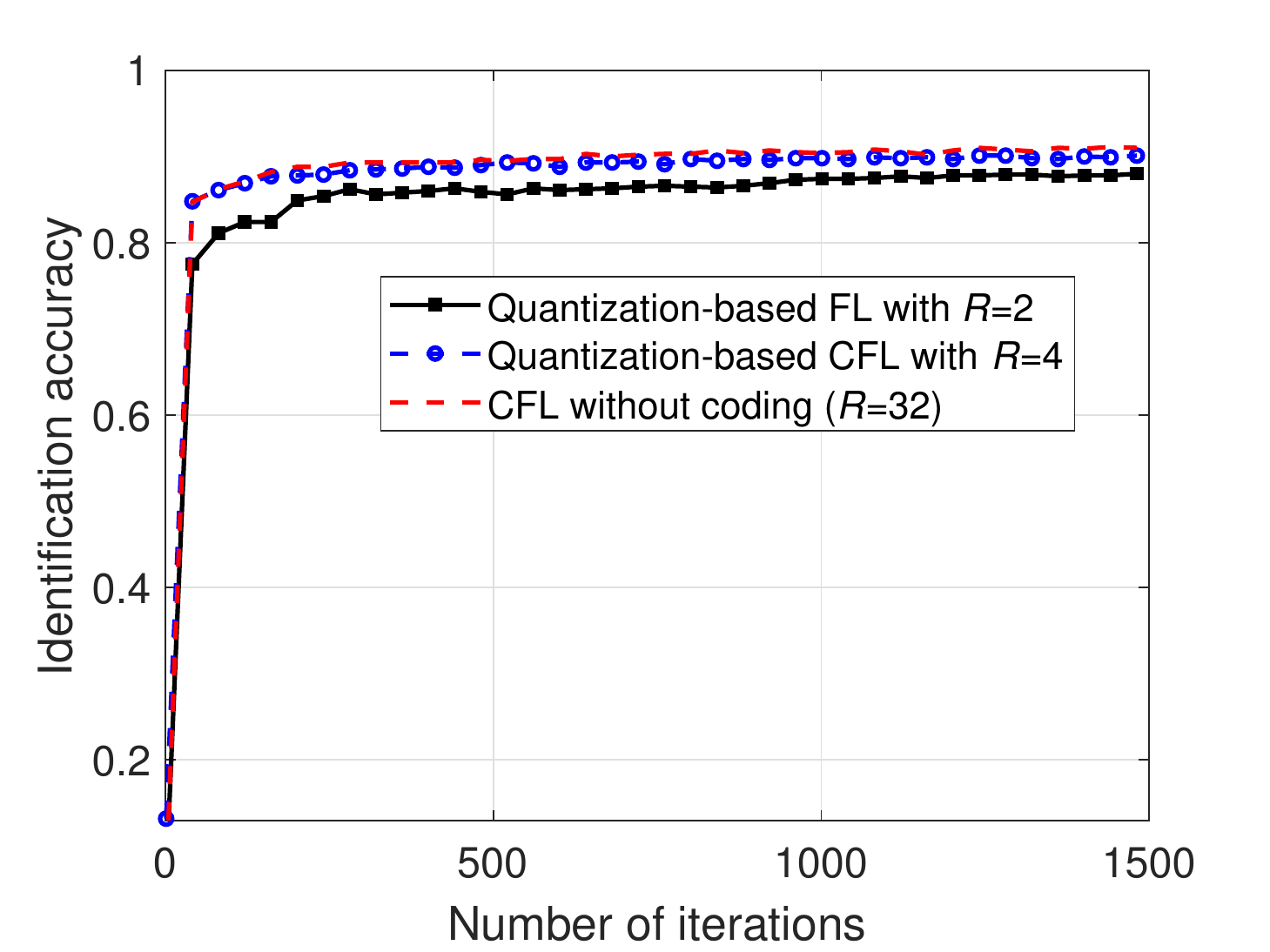}
\label{fig4b}} }
  \caption{Simulation scenario and result for source coding based FL.}\label{fig4}
\vspace{-0.3cm}
\end{figure*}

\subsection{Coding} 
During the CFL training process, source coding, channel coding, and gradient coding can be used to improve the FL performance. Source coding is used to compress the high-dimensional FL model parameters so that they can be represented by a small number of bits hence reducing the FL parameter transmission delay. Channel coding is used to protect the transmitted FL model parameters 
against the wireless noise and interference thus improving packet error rates and CFL reliability. Gradient coding is used to encode the GD parameters of machine learning algorithms so as to improve the ML performance. 

In this regard, a quantization-based source coding method was proposed by \cite{9054168} for reducing the data size of local FL models that are transmitted over wireless links. The coding and decoding procedure is shown in Fig. \ref{fig4a}. Here, we use the quantization-based coding method in \cite{9054168} for CFL. We implement a CFL algorithm for handwritten digit identification. All simulation settings are similar to the settings in \cite{9054168}.

 Fig. \ref{fig4b} shows how the accuracy of a handwritten digit identification learning task changes with the number of iterations. In Fig. \ref{fig4b},
 CFL uses $R$ bits to represent an element of the local FL model vector.
 Fig. \ref{fig4b} shows that, the quantization-based CFL algorithm with $R=4$ can almostly  achieve the same performance compared to the CFL algorithm without coding. Since the quantization-based CFL algorithm uses only $4$ bits to represent an element of the local FL model vector, the transmission delay of the quantization-based CFL algorithm will significantly decrease. From Fig. \ref{fig4b}, we can also see that the quantization-based CFL algorithm with $R=4$ can achieve better performance compared to the quantization-based CFL algorithm with $R=2$. This is because coding makes the local FL model after coding to be different from the FL model before coding. As the number of bits used to represent the local FL model decreases, the difference between the FL model after coding and the FL model before coding increases and thus affecting the identification accuracy. 


Obviously, source, channel, and gradient coding can significantly improve CFL performance. However, a number of research questions still exists:

 \begin{itemize}

\item \textbf{Heterogeneous source coding design}: In an IoT system, the wireless transmission link characteristics of each device will be different (e.g., different data rates). To efficiently use wireless resources for FL model transmission, each device may encode its local FL model using different number of bits or different coding techniques. This type of coding schemes is called heterogeneous source coding. For example, some devices can use 15 bits to represent their local FL models while another can use 7 bits to represent its local FL model.
Heterogeneous source coding can significantly reduce the coding energy consumption and decrease the loss function value. However, in CFL, a device must transmit its local FL model to multiple devices. Therefore, one must determine the number of local FL models that each device must encode and the number of bits used to encode the corresponding local FL models. For example, if a given device must transmit its local FL model to three devices, then this device can encode a local FL model and transmit it to three devices. Also, the device can encode two or three local FL models with different number of bits and then transmit them to these three devices.



\item \textbf{Gradient coding for avoiding stragglers}: Due to limited wireless resources, an IoT system has devices with extremely high transmission delay or computational delay. Such devices (called stragglers) may not be able to complete the local FL model transmission within the time duration required by the system. If a network has a large number of stragglers, the number of devices that can perform CFL will significantly decrease. Therefore, there is a need to design gradient coding schemes for addressing the problem of stragglers. However, traditional gradient coding methods 
require devices to share their dataset with other devices so as to remove stragglers and hence, they cannot be used for CFL since CFL does not allow devices to share their data. Hence, one must investigate a novel gradient coding scheme without data sharing.  
\end{itemize}

\section{Conclusion}\label{se:conclusion}
This article proposed a novel wireless FL framework, called collaborative FL and introduced the challenges and opportunities of using wireless communication techniques for optimizing CFL performance. 
The introduced wireless techniques provide guidance for reliably deploying CFL across edge IoT devices. The discussed research opportunities identify important open problems that must be considered when designing and deploying CFL for IoT systems. We expect that the proposed CFL framework will fundamentally change the original FL
architecture allowing it to be deployed for several future applications such as mobile keyboard prediction, IoT device identification and monitoring, and extreme event detection for autonomous vehicles.

\bibliographystyle{IEEEbib}
\bibliography{references1}
\footnotesize
\vspace{0.5cm}
\noindent {\bf Mingzhe Chen} (S'15, M'19) is currently a Post-Doctoral Researcher at the Electrical Engineering Department, Princeton University and at the Chinese University of Hong Kong, Shenzhen, China. His research interests include machine learning, virtual reality, unmanned aerial vehicles, wireless networks, and caching.

\vspace{0.2cm} 
 \noindent {\bf H. Vincent Poor} (S'72, M'77, SM'82, F'87) is
the Michael Henry Strater University Professor of Electrical Engineering. His research
interests are in the areas of information theory and signal processing, and their applications in wireless networks and related
fields such as energy systems and social networks.  
 
 \vspace{0.2cm}
\noindent {\bf Walid Saad} (S'07, M'10, SM'15, F'19) is a Professor at the Department of Electrical and Computer
Engineering at Virginia Tech. His research interests include wireless networks, machine learning, game theory, cybersecurity, unmanned aerial vehicles, and
cyber-physical systems.

 \vspace{0.2cm}
\noindent {\bf Shuguang Cui} (S'99, M'05, SM'12, F'14) is the Chair Professor of the Chinese University of Hong Kong at Shenzhen and the Vice Director at Shenzhen Research Institute of Big Data. His current research interests
focus on data driven large-scale system control and resource
management, large dataset analysis, IoT system design, energy-harvesting-based communication system design, and cognitive network optimization.
 
\end{document}